# Combined Orbital and Thermal Evolution of Oort Cloud Comets

Adam Parhi[1]⋆ and Dina Prialnik[1] †

[1]*Department of Geosciences, Tel Aviv University, Tel Aviv, Israel*

**ABSTRACT**
We present a fully integrated model of cometary evolution that couples thermal and compositional processes with dynamical processes continuously, from formation to present-day activity. The combined code takes into account changes in orbital parameters that define the heliocentric distance as a function of time, which is fed into the thermal/compositional evolution code. The latter includes a set of volatile species, gas flow through the porous interior, crystallization of amorphous ice, sublimation and refreezing of volatiles in the pores. We follow the evolution of three models, with radii of 2, 10 and 50 km for 4.6 Gyr, through different dynamical epochs, starting in the vicinity of Neptune, moving to the Oort cloud (OC) and after a long sojourn there, back inward to the planetary region. The initial composition includes a mixture of CO, $CO_2$ ices, amorphous water ice with trapped CO and $CO_2$, and dust. We find that the CO ice is completely depleted in the small object, but preserved in the larger ones from a depth of 500 m to the center, while the $CO_2$ and the amorphous ice are entirely preserved. Of crucial importance is the change in CO abundance profiles during the cooling phase, as the objects migrate to the OC. Upon return from the OC, the activity is driven by CO sublimation at large heliocentric distances (up to 50 au), by $CO_2$ inward of 13 au and by gas released from crystallizing amorphous ice at ∼ 7 au. We test the effect of radioactive heating by long-lived isotopes and find that it is negligible. Considering subsolar temperatures and limited active areas, we show that $CO_2$ production rates can exceed the detection limit as far out as 25 au.

## 1 INTRODUCTION

Comets are considered among the most primitive objects in the Solar system because they have spent most of their lives in a cold environment away from the Sun, where it is expected that they have maintained their original composition. As a result, they may contain significant information about the formation of early solar system planetesimals, as well as early primordial accretion processes (Bockelée-Morvan 2011). Beside water ice, which is the main cometary volatile, the most abundant volatiles are CO and $CO_2$, but other species, including $N_2$, $H_2S$, $CH_3OH$, $H_2CO$, $NH_3$, HCN, $CH_4$, and $S_2$, have also been detected in the coma and tails of comets (e.g. Disanti et al. 2002; Calmonte et al. 2016; Opitom et al. 2019; Kawakita et al. 2001; Cazaux et al. 2022). Refractory material, which accounts for at least half of the mass of the comet nucleus, is made up mainly of silicate minerals, dust, and organic compounds, such as hydrocarbons and tholins (Poch et al. 2016).

It is commonly assumed that cometary water ice is initially amorphous (Jenniskens & Blake 1996) and transforms into crystalline form in an exothermic, irreversible, and highly temperature-dependent process. The amount of latent heat released has been recently determined by Kushwaha et al. (2025). According to laboratory studies, amorphous ice can trap significant amounts of gas, most of which escapes when the ice crystallizes (Bar-Nun et al. 1987). The crystallization of amorphous ice has been considered as the mechanism responsible for ubiquitous outbursts observed in comets (see Prialnik & Jewitt 2024).

Cometary activity is generally driven by the sublimation of water ice close to the Sun, but comets are also found to be active at large heliocentric distances. Recent notable examples include C/2014 UN271 (Bernardinelli–Bernstein) – an Oort Cloud (OC) comet that was already active at 23.8 au (Farnham et al. 2021); C/2017 K2 (Pan-STARRS), detected active at 35 au (Jewitt et al. 2021); and C/2010 U3 (Boattini) (Milewski et al. 2024), another OC comet found active at 25.8 au pre-perihelion (Hui et al. 2019). More recently, C/2019 E3 (ATLAS; Hui et al. 2024) was observed to be active beyond 20 au, and the long-period comet C/2024 E1 (Wierzchos) exhibited $CO_2$-driven activity at a heliocentric distance of 7 au (Snodgrass et al. 2025). The immediate conclusion has been that at such large distances, where temperatures are too low for the sublimation of water ice, cometary activity must be driven by the sublimation of more volatile species.

These observations challenge traditional models of cometary activity, raising the question of whether supervolatile ices can survive the early evolutionary stages of comet nuclei. For example, in a previous study, we have found that a comet that spends most of its life in the Kuiper Belt (KB) should lose all its free volatiles (Parhi & Prialnik 2023), other than water ice. However, comets that have become active at large distances are long-period comets originating in the OC rather than the KB. As such, at some point in their life they traveled far enough from the planets into low-temperature environments that may have allowed them to preserve their volatiles.

⋆ E-mail:adamparhi@tauex.tau.ac.il
† E-mail:dinak@tauex.tau.ac.il

The purpose of the present paper is to follow the long-term evolution of OC comets from the early stages spent in the planetary region, through migration to the OC and back. There have already been attempts to understand the long-term thermal processing of comet nuclei using numerical models or theoretical estimates. The simplest approach is to consider the saturated vapor pressure that controls the sublimation rate as a function of temperature, for the most common volatiles detected in comets. Since the temperature is related to the heliocentric distance, this gives an idea of the time needed to lose a volatile species at a given distance from the Sun (Lisse et al. 2021). However, this remains an approximate estimate, since it presumes that all volatile constituents of the comet body are uniformly exposed to solar radiation, without considering the internal composition and structure of the comet nucleus.

Other studies (Gkotsinas et al. 2022; Guilbert-Lepoutre et al. 2023) combined dynamical and thermal evolution of distant comets, but ignored energy exchange in phase transitions, using evolutionary isotherms to derive the depths of volatile ices. The authors explicitly acknowledged that temperature profiles alone are insufficient —- and potentially misleading —- indicators of internal compositional evolution. More recently, Gkotsinas et al. (2024) considered the combined thermal and dynamical evolution of planetesimals scattered in the OC, accounting for sublimation of volatiles and the accompanying energy absorption by means of an approximate, averaged (asynchronous) method (see Schörghofer & Hsieh 2018) that does not account, however, for gas flow or recondensation. The evolution of planetesimals during the early history of the solar system was also investigated by Davidsson (2021), by means of a complex evolution code (NIMBUS), which includes water vapor flow through pores, but considers CO and $CO_2$ sublimation and loss in a simplified form. The conclusions of these studies are controversial: while Davidsson (2021) finds that CO ice is completely lost, Gkotsinas et al. (2024) conclude that CO ice may survive in the interior of comets originating in the OC.

Prompted by the puzzling modeling results on the one hand and by the recent observations on the other, we undertake to address the long-term evolution problem in its entirety, taking account of orbital dynamics, together with thermal and detailed compositional evolution for water, CO and $CO_2$ ices. We will show that recondensation and flow of volatiles play a crucial role in explaining the present activity of OC comets. In Section 2 we describe the dynamical evolution path adopted; in Section 3 we briefly describe the thermophysical model employed; in the following Section 4 we present the results of the numerical simulations; finally, in Section 5 we discuss the results in light of observations and summarize our conclusions.

## 2 ORBITAL EVOLUTION

Following the formation of the planets, the solar system contained a large population of planetesimals distributed in the outer regions of the protoplanetary disc. Gravitational interactions with the giant planets perturbed the orbits of these planetesimals, scattering them both inward and outward. Those scattered outward had their semi-major axes increased to hundreds and thousands of au in very eccentric orbits. While many of the planetesimals were ejected from the solar system, a considerable fraction became gravitationally trapped at very large distances, forming the OC.

Detailed N-body hydrodynamical simulations (Dones et al. 2004; Brasser et al. 2007; Kaib & Quinn 2008; Brasser & Morbidelli 2013) have shown that the main feeding zone of OC comets was roughly 15-35 au from the sun, peaking around 20-30 au, which corresponds to the region of Uranus and Neptune. The formation of the OC began after the formation of the giant planets, mainly within 10-50 Myr and continued for another 100 Myr (Dones et al. 2015; Nesvorný et al. 2017; Portegies Zwart et al. 2021). If the motion of these bodies were affected by planetary gravitational forces alone, they would ultimately be ejected from the solar system. Therefore, an additional stabilizing mechanism must have existed to keep them in distant orbits. Possible candidates for this role include perturbations from passing stars (e.g., Fernández 1997; Brasser et al. 2006) and the galactic tidal field (e.g., Harrington 1985; Fouchard 2004), the latter being considered to provide a more significant long-term influence on the evolution of the orbits, inducing coupled oscillations between orbital eccentricity and inclination (Heisler & Tremaine 1986). Higuchi et al. (2007) showed that for comets with semi-major axes around 10,000 au or greater, these oscillations can persist for timescales comparable to the age of the solar system.

Occasionally, galactic tidal perturbations may cause the perihelion distance to be driven inward from thousands to a few au. Passing stars that come within a few thousand au from the sun can also cause significant orbit perturbations (Kaib & Quinn 2009; Fouchard et al. 2011; Rickman 2014). Only a few of these comets return to the inner planetary region. These "new comets" are the objects of interest in the present study.

To follow the internal long-term evolution of OC comets from their formation to present time, we need to provide a continuous function for the time dependence of the heliocentric distance $d(t)$. Adopting the main features of the dynamical evolution of planetesimals just described, we constructed a generic representative evolution path spanning 4.6 Gyr, which can be viewed as a continuous series of eccentric orbits with varying perihelion distance $q$ and semi-major axis $a$. It may be roughly divided into four phases. In the first phase, the comet starts with a perihelion distance of approximately 30 au, and the semi-major axis increases gradually, while the perihelion remains constant, allowing the comet to migrate outward until its aphelion reaches the lower boundary of the OC at around 2,000 au after approximately 30 Myr. In the second phase, the orbital eccentricity decreases, causing the perihelion to increase beyond 100 au. This phase is short, lasting about 3 Myr. In the third and longest evolutionary phase, the comet resides in the OC; the semi-major axis stabilizes at ∼10,000 au, while the eccentricity starts to oscillate and gradually decreases, reaching a minimum value of 0.6, corresponding to a perihelion distance of about 4,000 au, over a span of 4.5 Gyr. In the final, fourth phase, the eccentricity increases to a sufficiently high value of approximately 0.9995, meaning that the perihelion distance decreases to about 5 au, thereby bringing the comet back into the inner Solar System. The progression through these orbital stages is illustrated in Fig.1, which shows the variation of perihelion and aphelion distances of the orbits throughout evolution, marking the key dynamical transitions; the total number of orbits exceeds 180,000.

Long-term evolution calculations are extremely time consuming. One reason is that following eccentric orbits with high resolution requires small time steps. In order to simplify calculations involving elliptical orbits, an averaging method is often employed to replace them by circular orbits. For example, Prialnik & Rosenberg (2009) adopted circular orbits that receive the same total solar energy over an orbital period. Gkotsinas et al. (2023) have addressed this problem in detail, comparing various methods and analyzing their strengths and weaknesses. The main drawback is that comets may attain very high temperatures at perihelion, potentially reaching thresholds sufficient to induce the sublimation of volatiles, which could be missed by an average temperature of a circular orbit.

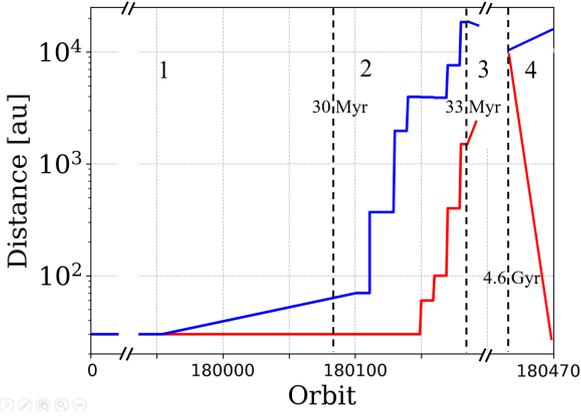

**Figure 1.** The dynamical history presented by the changing perihelion (red) and aphelion (blue) distances with orbit number for the 4.6 Gyr of evolution, marking the four dynamical phases: (1) early evolution in the planetary region, (2) outward migration toward the OC, (3) long-term residence in the OC, and (4) return to the inner Solar System.

In this paper, we refrain from such approximations and follow the coupled dynamical and thermal/compositional evolution along the eccentric orbits accurately, capturing the effects of orbital evolution on thermal processes and volatile activity. We consider models of three different sizes — 2, 10, and 50 km — adopting the same dynamical path for all three.

## 3 THERMAL AND COMPOSITIONAL EVOLUTION

The evolution model used in this study, based on a spherically symmetric configuration, is described in detail in Prialnik (1992); Prialnik et al. (2004). We assume a composition of dust (refractory material), $H_2O$ (amorphous and crystalline), and $N$ other volatile species, in a highly porous structure. The partial densities of volatile species are indicated by $\rho_{s,n}$ for the solid phase and $\rho_{v,n}$ for the vapor phase ($1 \le n \le N$), $\rho_a$ and $\rho_c$ are the partial densities of amorphous and crystalline water ice, respectively, $\rho_v$ is the partial density of water vapor, and $\rho_d$ is the partial density of dust. Thus,

$$\rho = \sum_n (\rho_{s,n} + \rho_{v,n}) + \rho_a + \rho_c + \rho_v + \rho_d \quad (1)$$

where $n$ runs over all volatile species. The porosity is given by

$$\psi = 1 - \sum_n \rho_{s,n}/\varrho_n - (\rho_a + \rho_c)/\varrho_{H_2O} - \rho_d/\varrho_d, \quad (2)$$

where $\varrho$ denotes the characteristic density of the nonporous solid phase (for example, $\varrho_{H_2O} = 917$ kg m$^{-3}$). The energy per unit volume is given by

$$\rho u = \sum_n (\rho_{s,n} u_{s,n} + \rho_{v,n} u_{v,n}) + (\rho_a + \rho_c) u_{H_2O} + \rho_v u_v + \rho_d u_d, \quad (3)$$

where the specific energies $u$ are functions of temperature. Let **F** denote the heat flux, $\mathbf{F} = -K(T)\nabla T$, where $K(T)$ is the thermal conductivity. The volume rate of sublimation for each volatile species $q_n$ is given by $q = SZ(T)$, where $S$ is the surface to volume ratio characterizing the porous medium and $Z(T) = \sqrt{m/2\pi kT}[A\exp(-B/T) - P]$ is the sublimation rate (Mekler et al. 1990). Here $m$ is the molecular mass, $A$ and $B$ are the coefficients of the saturated vapor pressure in the Clausius-Clapeyron approximation, and $P$ is the gas pressure. The gas densities in the cases considered are very low, and hence the flow is in the Knudsen regime, thus the gas (vapor) fluxes $\mathbf{J}_n$ are given by $\mathbf{J}_n = -\phi_n \nabla(P_n/\sqrt{T})$, where the permeability $\phi$ depends on porosity, pore size and tortuosity.

The temperature-dependent rate of crystallization (s$^{-1}$) was derived experimentally by Schmitt et al. (1989) in the form

$$\lambda(T) = c\exp(-b/T), \quad (4)$$

with $c = 1.05 \times 10^{13}$ s$^{-1}$ and $b = 5370$ K. Volatile species may be trapped in amorphous ice in mass fractions $f_n$ relative to the ice and released upon crystallization. We denote the entire fraction of trapped volatiles by $f = \sum_n f_n$. The heat released in the crystallization process $H_a$ has recently been experimentally re-evaluated by Kushwaha et al. (2025).

The evolution of the body's structure is described by a set of conservation laws. The mass conservation equations are:

$$\frac{d\rho_a}{dt} = -\lambda\rho_a, \quad \frac{d\rho_c}{dt} = \lambda\rho_a(1-f) - q_v, \quad \frac{\partial\rho_v}{\partial t} + \nabla \cdot \mathbf{J}_v = q_v, \quad (5)$$

$$\frac{d\rho_{s,n}}{\partial t} = -q_n, \quad (6)$$

$$\frac{\partial \rho_{v,n}}{\partial t} + \nabla \cdot \mathbf{J}_n = q_n + \lambda \rho_a f_n \quad (7)$$

with $n = 1, 2$ in our case. The energy conservation law yields:

$$\frac{\partial}{\partial t}(\rho u) + \nabla \cdot \left(\mathbf{F} + \sum_n u_{v,n}\mathbf{J}_n\right) = -\sum_\alpha q_n H_n + \lambda\rho_a H_a + \dot{Q}. \quad (8)$$

The second term on the left-hand side represents the transfer of heat by conduction and advection by flowing volatiles, if present, and the terms on the right-hand side include the rates of absorption/release of latent heat by sublimation/condensation in pores and radioactive energy release $\dot{Q}$. The implicit assumption is that all the components of the body are in local thermodynamic equilibrium and hence a unique local temperature may be defined. The set of non-linear time-dependent second-order partial differential equations (6)-(8) is solved numerically on a sphere, using an implicit iterative scheme. To ensure global energy conservation and mass conservation for all species throughout the entire evolution, very stringent convergence accuracy criteria were imposed.

The solution requires initial and boundary conditions. The inner boundary conditions are vanishing fluxes of mass and heat at the center of the sphere. At the surface, the gas pressures are assumed to vanish. The surface heat flux $F_s$ is given by

$$\frac{(1-A)L_\odot(t)}{16\pi d(t)^2} = \epsilon\sigma T_s^4 + Z(T_s)H - F_s, \quad (9)$$

where $L_\odot$ is the solar luminosity and $d$ is the heliocentric distance. It is assumed that the solar energy is spread uniformly over the surface (fast-rotator approximation). This assumption is valid for modeling long-term evolution, where the interior structure, below skin depth, is considered. When surface or sub-surface processes are involved, the solar zenith angle plays a role. We return to this point in Section 5.2.

We take into account the evolution of the solar luminosity during the early stages before the Sun settles on the main sequence according to the solar model computed by Kovetz et al. (2009), as shown in Fig. 2. Although this phase is relatively brief, it is bound to affect the sublimation of volatile ices and provide more realistic conditions under which young comets have evolved.

The initial homogeneous conditions and the physical parameters adopted for all models are summarized in Table 1.

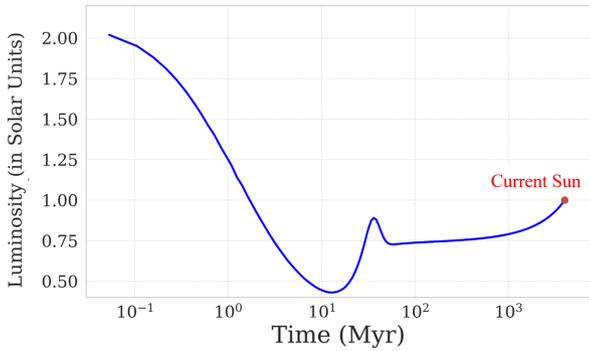

**Figure 2.** Evolution of the solar luminosity, as derived by Kovetz et al. (2009), until the time when the Sun reaches the present luminosity (marked), which remains constant until present time.

**Table 1.** Initial conditions and physical parameters.

| Parameter | Value |
| --- | --- |
| Radius | 2, 10, 50 km |
| Initial uniform temperature | 16 K |
| Initial dust mass fraction | 0.5 |
| Density | 500 kg m$^{-3}$ |
| Porosity | 0.66 |
| Albedo | 0.06 |
| CO ice relative to water ice | 10% |
| CO$_2$ ice relative to water ice | 10% |
| Trapped CO fraction ($f_{CO}$) | 5% |
| Trapped CO$_2$ fraction ($f_{CO_2}$) | 5% |
| Heat capacity of ice | 7.49T+90 J kg$^{-1}$ K$^{-1}$ |
| Heat capacity of dust | 130 J kg$^{-1}$ K$^{-1}$ |
| Thermal conductivity (at 20 K) | 0.0035 W m$^{-1}$ K$^{-1}$ |
| Thermal inertia (at 20 K) | 20 J m$^{-2}$ K$^{-1}$ s$^{-1/2}$ |

## 4 RESULTS OF EVOLUTION SIMULATIONS

The thermal and compositional evolution of the comet models along the path shown in Fig. 1 occurs mainly during the short phases: in the early phase 1, which is spent relatively close to the Sun, in phase 2, when the comet recedes to large distances and the surface cools more rapidly than the interior, and again towards the end of phase 4, when the comet returns from the OC to the solar neighborhood. During the long intervening period of time spent in the OC—phase 3—the comet is dormant.

The most important question that we aim to answer is, when would the returning comet become sufficiently active to be detected and which process would trigger such activity. A priori, there are three possible sources of activity: sublimation of CO, as representative of hypervolatile species, sublimation of CO$_2$, representative of less volatile ices — both abundantly detected in most comets (Bockelée-Morvan & Biver 2017) — and crystallization of amorphous ice, which is accompanied by the release of occluded gases (Bar-Nun et al. 1987). Sublimation of water ice is less relevant in this respect, as we are looking for distant activity, which characterizes many of the observed long-period comets.

### 4.1 Phase 1 - Early evolution in the planetary region

Our three comet models, with radii of 2, 10, and 50 km, begin their orbital journey in the inner solar system at a distance of 30 au. We refer to this initial orbital stage as the planetary region phase, during which the models maintain a nearly constant perihelion distance of about 30 au. Over time, the semi-major axis gradually increases and by the end of this phase, after about 30 Myr, the aphelion distance reaches roughly 100 au. Fig. 3 shows the temperature and CO abundance profiles throughout this phase. The CO sublimation front follows the advancing heat diffusion depth. The effect of cometary size is clearly seen: the small comet is heated more efficiently, resulting in a complete depletion of CO by the end of this phase. In the larger models, the CO depletion depth $\Delta R$ extends to about 500 m, slightly larger for the model with a 10 km radius, as compared to the 50 km radius one. The weak dependence on radius is expected when $\Delta R \ll R$, because then $\dot{R}_{CO} \approx L_\odot/(16\pi d^2 \rho_{CO} H_{CO})$, where $R_{CO} = R - \Delta R$ and $H_{CO}$ is the latent heat of CO sublimation. There is a typical peak in CO abundance at the sublimation front that cannot be discerned in the figure (see Fig. 6). The CO$_2$ ice, on the other hand, is almost completely preserved in all models, as temperatures remain lower than the sublimation temperature of CO$_2$; only a ~10 cm deep outer layer is depleted of CO$_2$ when the comet returns from the OC. The amorphous ice is preserved throughout.

Here we draw attention to the importance of rigorously taking into account the processes of internal sublimation of volatile ices and vapor flow through the porous interior. To this purpose, we have simulated the evolution of two identical models, adopting the same dynamical path for a period of 1 Myr, one with the full thermal model and the other, with no volatiles except water ice, to determine the depth corresponding to the CO sublimation temperature. Comparing it with the temperature profile and the actual location of the CO front as obtained for the model that includes the CO ice and vapor, the difference is obvious, as shown in Fig.4. The reason for the discrepancy is first, that in the correct calculation, part of the solar energy goes into the latent heat of CO sublimation, and secondly, the advance of the sublimation front depends on the thermal conductivity and diffusivity of the porous medium.

### 4.2 Phase 2 - Orbital expansion towards the OC

To summarize, by the end of phase 1, CO ice is retained in large comets, but it is depleted down to depths of order 500 meters. If this were to remain the case until the comets return from the OC as "new comets", the CO ice would lie far too deep to be activated by solar heat. This can be shown by a simple analytical estimate. Adopting an arbitrary orbit with a very large semi-major axis, say $a = 2000$ au, and a perihelion distance of 5 au, we can combine the relations $d(t)$ and the local equilibrium temperature $T(d)$ to derive the surface temperature as function of time along the orbit $T(t)$, as shown in the upper panel of Fig.5. Assuming that heat diffusion to the interior becomes significant when $T(t)$ exceeds the CO sublimation temperature (roughly 25 K), we can estimate the heat diffusion depth from that point in time onward by $Z_{diff} = \sqrt{2\tau_{diff}\kappa}$, where $\kappa$ is the thermal diffusivity and $\tau_{diff}$ the diffusion time, and derive $Z_{diff}(d)$ as along the orbit, as shown in the lower panel of Fig.5.

The conclusion is that the depth of heat penetration does not exceed a few tens of meters, much less than the depth of CO ice resulting from phase 1. This apparent puzzle will be solved as we follow the move of the comet to colder regions of the solar system during phase 2 of the dynamical evolution.

At the beginning of phase 2, near the CO ice front, the temperature is still high enough for sublimation to occur; however, the surface layers become too cold for the vapor to escape - instead it refreezes. The CO ice profile and the temperature profile at the beginning of this phase are shown in Fig. 6.

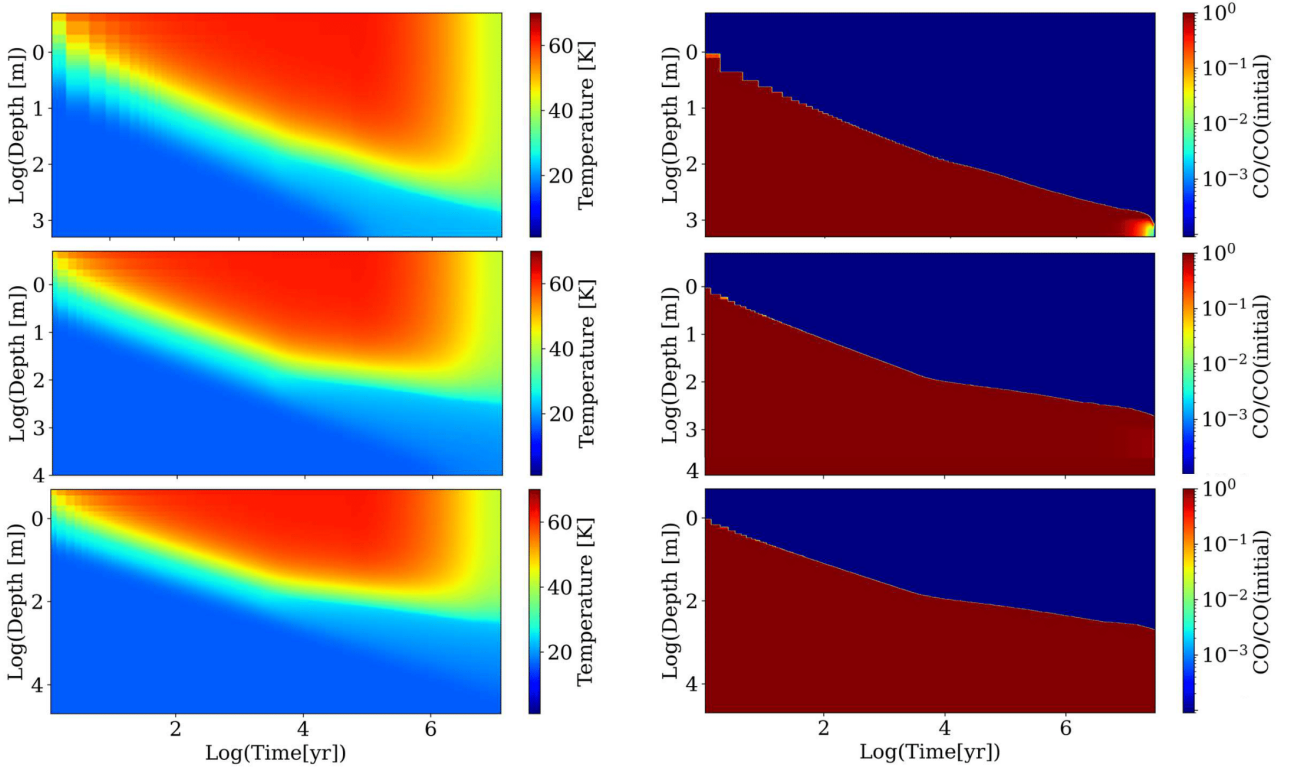

**Figure 3.** Temperature profiles (*left panels*) and CO abundance profiles (*right panels*) for comets with radii of 2 km (*top*), 10 km (*middle*), and 50 km (*bottom*) during phase 1 near 30 au. The smallest object becomes entirely depleted of CO, while in larger bodies the CO sublimation front retreats to depths of several hundred meters, illustrating the strong dependence of volatile retention on comet size.

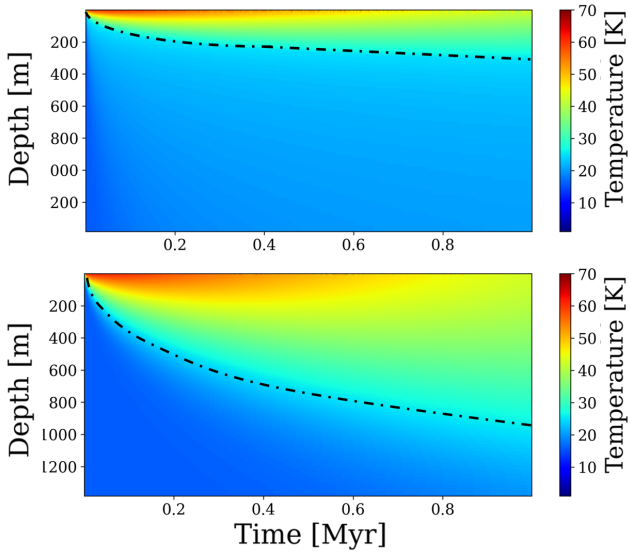

**Figure 4.** Temperature profiles for a 10 km comet model during the first 1 Myr of evolution spent near Neptune, computed with and without gas physics. *Upper panel*—full model including gas: the dashed line represents the true CO depletion depth. *Lower panel*—thermal model without gas: the dashed line follows the CO sublimation temperature, taken to represent the CO depletion depth. The striking difference shows that gas-related processes must be included along with thermal processes.

The resulting scenario is that of coupled heat and mass transfer, with feedback due to phase transitions. The system of coupled non-linear partial differential equations that describes this scenario constitutes a well-known problem (encountered, for example, in the study of permafrost), whose solution is unstable, resulting in oscillations in both time and space (Adams & Brown 1990; Hansen & Foslien 2015). The numerical solution requires fine zoning and small time steps (e.g. Brondex et al. 2023). The evolution of the CO profile during this stage is shown in Fig. 7.

Thus, at the end of phase 2, the CO profile is oscillatory, as illustrated in Fig. 8 for the outer 200 m of the R=50 km comet model; it has a pronounced peak at a depth of a few tens of meters, which we have identified as the heat diffusion depth on the return orbit from the OC. It is this ice that will start CO sublimation-driven activity later on (see Section 4.4). This effect is missed by calculations that ignore gas flow and recondensation (e.g., Gkotsinas et al. 2024).

### 4.3 Phase 3 - Stable evolution in the OC

During the third orbital phase, the stable OC stage, the comet remains at large heliocentric distances for nearly 4 Gyr and cools continuously throughout, as shown in Fig.9. As temperatures remain below 25 K—the sublimation threshold of CO—everywhere, the volatiles of the comet remain stable and preserved in a frozen state, as they were at the end of the previous phase. The solar input energy is very low during this long phase, but one should consider the energy provided by long-lived radioactive nuclei—$^{40}$K, $^{232}$Th, $^{235}$U and $^{238}$U—which is negligible for small objects in the planetary region, but becomes comparable and may exceed the solar heating rate at OC distances.

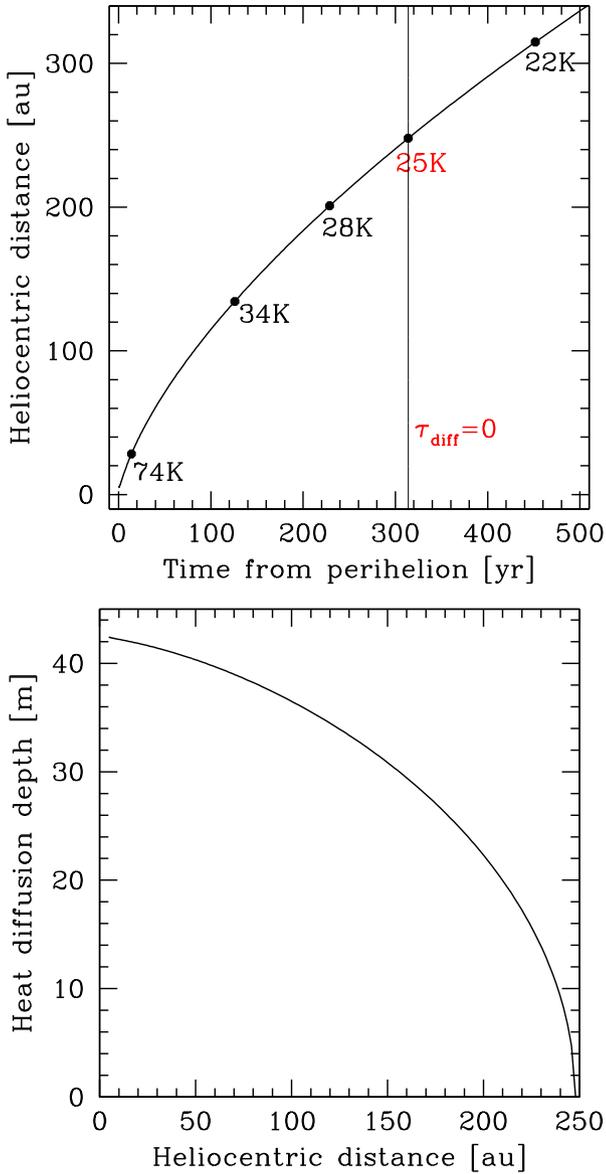

**Figure 5.** *Upper panel*: Fragment of the cometary orbit from the OC with the corresponding surface temperature values marked. The onset of surface CO sublimation (∼ 25 K) is taken to represent the zero point of the inward heat diffusion time. *Lower panel*: The heat diffusion depth corresponding to the diffusion time $\tau_{\rm diff}$ as a function of heliocentric distance. At perihelion the heat penetration depth does not exceed a few tens of meters.

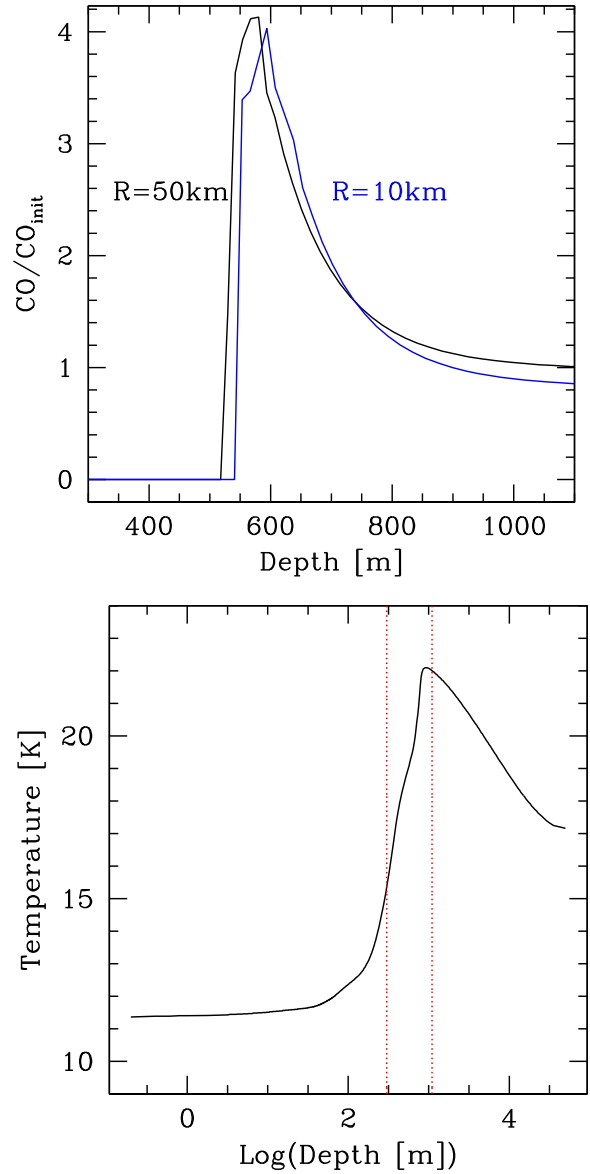

**Figure 6.** *Upper panel*: Normalized CO ice profile at at the beginning of phase 2, when the comet migrates outward toward the OC: for the 50 km model (black) and for the 10 km model (blue), which is similar. *Lower panel*: The temperature profile throughout the 50 km model at the same point in time, where the region corresponding to the upper panel is marked. Vapor from the CO front as well as heat from the corresponding temperature peak flow to the surface.

To test this effect, we have repeated the simulation for the 50 km radius model including long-lived radioactive species in meteoritic abundances, as given by Davidsson (2021). The main effect is slightly elevated temperatures, as shown in Fig.9, very little sublimation of CO ice in the central region, but also a higher cooling flux by a factor of 2, corresponding to the difference in surface temperatures with and without radioactives at the same distance throughout this phase. The effect on the next evolutionary phase is minor, as we shall show shortly. The $CO_2$ ice, which lies close to the surface, is not affected at all, nor is the amorphous water ice.

This result is easily understood if we consider the maximal rate of radioactive energy supply, given by $\dot{Q}_{\rm max} = \sum X_{0,i} \rho_d H_i / \tau_i$, where $X_0$ is the mass fraction of a radiative species relative to dust, $H_0$ is the decay energy per unit mass, and $\tau_i$ is the decay timescale, with $i$ running over all species. The surface temperature $T_s$ that would be required to get rid of this energy together with the solar energy is obtained by

$$\sigma T_s = \left( \frac{L_\odot}{16\pi d^2} + \frac{\dot{Q}_{\rm max} R}{3} \right)^{1/4} \qquad (10)$$

and amounts to only ∼ 2 K above the local equilibrium temperature, as seen in Fig. 9. During this long phase, most of the radioactive energy is simply radiated away.

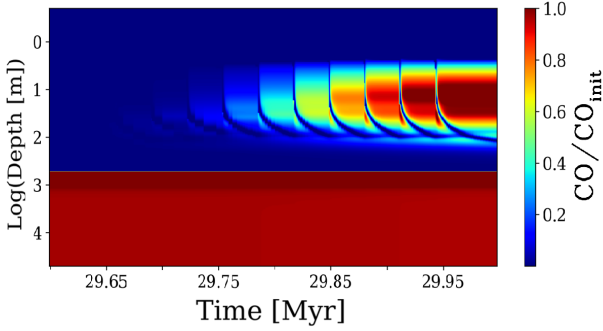

**Figure 7.** Evolution of the normalized CO abundance profile during phase 2, as the comet migrates outward toward the OC. The vapor produced at the CO ice front refreezes in the cooler outer layers, leading to an oscillatory abundance pattern. The thin lines crossing the profile indicate perihelion passages, during which the temporary rise in temperature enhances CO sublimation.

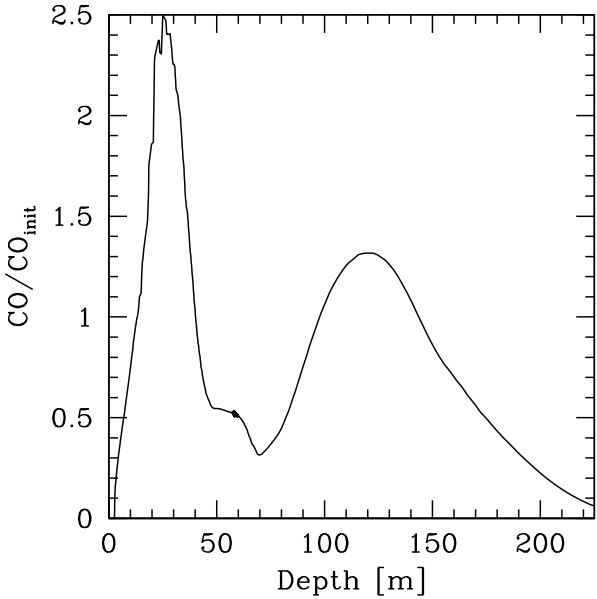

**Figure 8.** Normalized CO abundance profile at the end of phase 2 in the outer layer of the 50 km comet model. The distribution exhibits an oscillatory structure with a pronounced peak at a depth of several tens of meters. This frozen CO layer later becomes the main source of activity upon the comet's return from the OC.

### 4.4 Phase 4 - Return from the OC to the planetary region

In the next and final stage, the comets return to the inner solar system on a very eccentric orbit, appearing as new comets. At large heliocentric distances, they are expected to be activated by sublimation of CO and $CO_2$ ices. We assume that the detection limits of CO and $CO_2$ emission correspond to rates of $10^{26}$ molecules s$^{-1}$ and $2 \times 10^{25}$ molecules s$^{-1}$, respectively (Ye et al. 2023). We find that on the inbound leg of the orbit CO production by sublimation in the upper layers of the comet first exceeds the detection threshold at a heliocentric distance of ~50 au for the 10 km model and ~150 au for the 50 km model, as shown in the upper panel of Fig. 10. As the comet approaches the Sun, the surface temperature increases and $CO_2$ starts sublimating as well, reaching detection limit at about 13 au. We note that just below 10 au, the $CO_2$ production rate exceeds the CO pro-duction rate for the 10 km radius object. We also draw attention to the settling of $CO_2$ production after the initial peak; this is due to the recess of the sublimation front from the surface. The small object, which is devoid of CO, except for the fraction trapped in the amorphous ice, emits only $CO_2$, reaching detection limit between 7-8 au. The production rates scale as the surface areas of the comets.

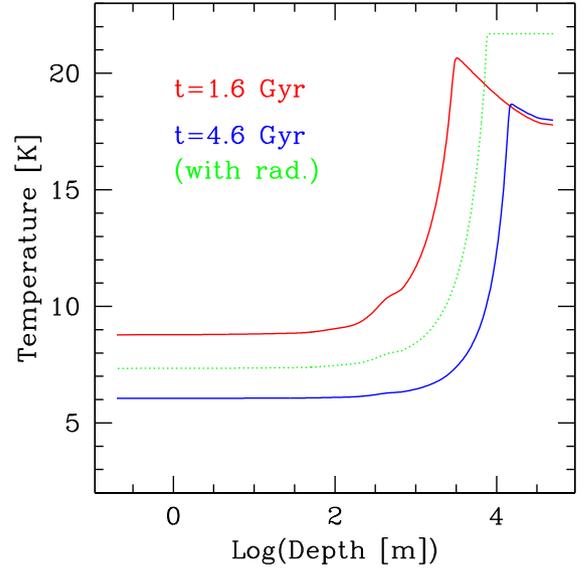

**Figure 9.** Temperature profiles throughout the 50 km model during phase 3: without radiogenic heating, at 1.6 Gyr (red) and at 4.6 Gyr (blue), and the corresponding profile including heating by long-lived radioactive isotopes at 4.6 Gyr (dotted green). In all cases, temperatures remain below the CO sublimation threshold (~25 K), ensuring retention of volatile ices. Radiogenic heating produces only a marginal temperature increase (≲ 2 K), indicating that radioactive decay has a negligible effect on the long-term thermal evolution of the comet within the OC.

The other source of activity at large distances is the release of occluded gases upon crystallization of amorphous ice, which has survived early evolution intact (see Section 4.1). In this case, volatiles are released in more or less the same ratios as they they have been trapped in the ice. The crystallization of amorphous ice sets in gradually around 8 au. In all cases, at a distance of about 6.5 au, the exponentially rising rate of crystallization becomes the main source of volatiles, exhibiting a sharp increase in the outgassing rates to a total production of $10^{28}$ molecules s$^{-1}$ and $10^{29}$ molecules s$^{-1}$ for the 10 km and 50 km models, respectively, for both CO and $CO_2$.

## 5 DISCUSSION AND CONCLUSIONS

### 5.1 Comparison with observations

We now focus on the results of phase 4 and emphasizing the fact that they are based on a single, albeit typical, dynamical path, we confront them with early detections of OC comets, keeping in mind the difficulty in determining production rates of volatiles at large distances. In many cases, these are derived from the measured dust production rates, assuming a dust/gas mass ratio $\chi$, e.g., $\chi = 5$ (Jewitt et al. 2019). Only recently has JWST enabled accurate determinations from spectral observations of distant comets (e.g., Snodgrass et al. 2025). In Table 2 we list the detection distances of several OC comets, CO and $CO_2$ production rates, where measured or estimated, and the corresponding references, and sizes, where known.

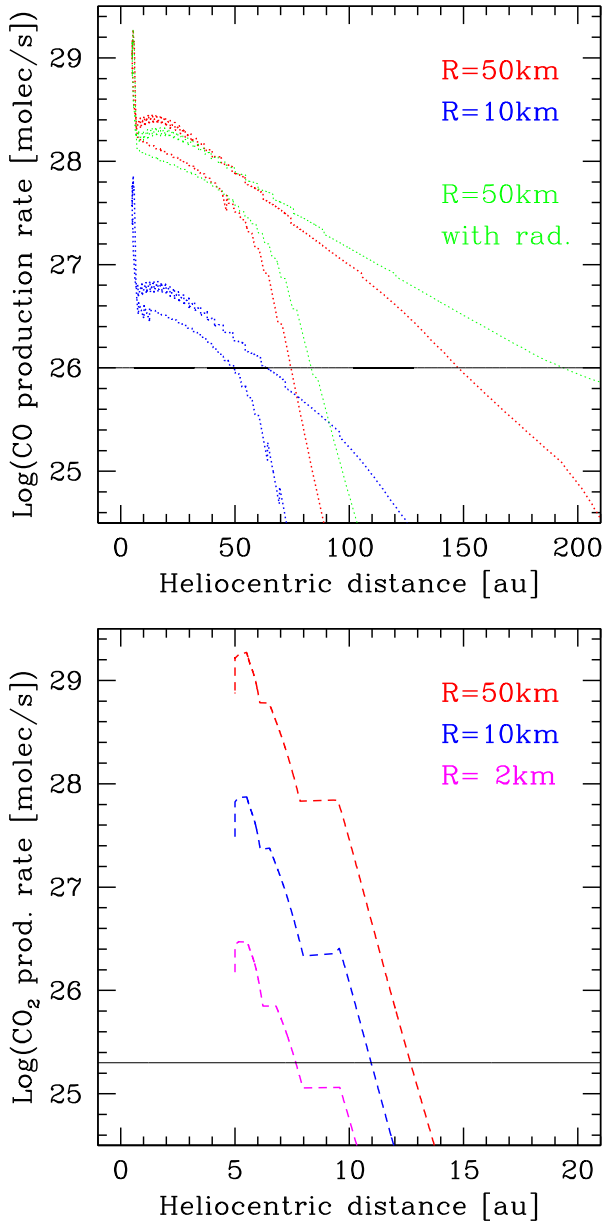

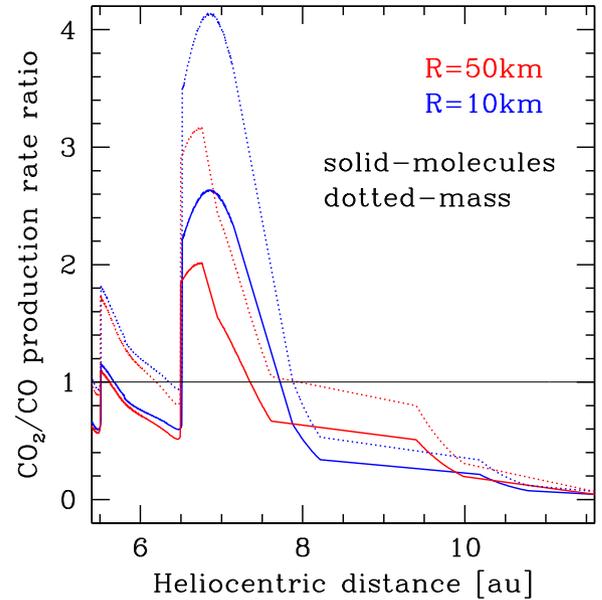

**Figure 11.** The ratio of $CO_2$ to CO production rates as function of heliocentric distance obtained for the 50 km (red) and 10 km (blue) models. The ratios are presented both in terms of number of molecules s$^{-1}$ and of mass s$^{-1}$.

large comet (R=50 km), in agreement with the radii derived from observations : ~ 70 km for C/2014 UN271, and an upper limit of 80 km for C/2017 K2. In fact, none of the observational results is in conflict with the results of our calculations.

According to the lower panel of Fig. 10, activity driven by $CO_2$ emission should start at ~ 13 au, while the release of trapped volatiles from crystallizing amorphous ice exceeds the other sources inward of 7 au. The ratio of $CO_2$ to CO production rates, both in terms of molecule numbers and of mass, is shown as a function of heliocentric distance in Fig. 11: the $CO_2$ production rate dominates CO production in a narrow range of heliocentric distances, in our case 6.5-8 au, the effect increasing with decreasing comet size.

The precise determination of CO2 production rate in comet C/2024 E1 and the lack of CO detection indicate that the comet may be small. We have found that the upper limit for the radius of an object that would become completely depleted of CO lies in the range $2 < R(km) < 10$. Of special interest is comet C/2014 B1, for which the $CO_2$ production rate is significantly higher than that of CO at the distance of 9.7 au. Our prediction would be that this is a relatively small comet, in agreement with the 6.4 km radius determination by Ivanova et al. (2023).

**Figure 10.** *Upper panel*: CO production rates as functions of heliocentric distance during the last orbit (phase 4) for the 50 km model (red) and the 10 km model (blue). The results for the 50 km model including radioactive heating are also shown (green). The horizontal line marks the CO detection limit (see text). *Lower panel*: $CO_2$ production rates as functions of heliocentric distance during the inbound leg of the last orbit (the outbound leg production rates are similar in this case) for all models, as marked. The black horizontal line marks the $CO_2$ detection limit.

As shown in the upper panel of Fig. 10, CO emission-driven activity should start around 50 au on the inbound leg of the orbit. Indeed, the farthest distance where activity was detected is 35 au (comet C/2017 K2) and we predict that it was driven by CO sublimation from the outer few tens of meters of—most probably—a large comet. An upper limit of $2 \times 10^{28}$ molecules s$^{-1}$ on the CO production rate was inferred for comet C/2014 UN271 at 24 au and this estimate is also compatible with our results. Around 16 au there are several estimates of CO production rate: $3.4 \times 10^{27}$ (C/2014 UN271) and $1 - 1.6 \times 10^{28}$ (C/2017 K2). They fit the model results for the

### 5.2 Theoretical upper limit for activity-detection distance

A few OC comets have shown activity triggered by $CO_2$ beyond the limit of 13 au (e.g. Meech et al. 2017; Milewski et al. 2024; Roth et al. 2025). Comet C/2017 K2, discovered by the Pan-STARRS survey (Wainscoat et al. 2017) and later identified in pre-discovery images (Jewitt et al. 2017; Hui et al. 2017) showed significant activity at a heliocentric distance of 23.4 au, with a reported $CO_2$ production rate of $2.56 \times 10^{26}$ molecules s$^{-1}$.

To account for such distant activity we should keep in mind that our comet model assumes a spherically symmetric configuration, where the solar energy is evenly distributed over the surface, resulting in a unique temperature, which averages diurnal temperature variations.

**Table 2.** Distant activity detected (estimated) for OC comets

| Comet | $d_{obs}$ [au] | CO [n/s] | $CO_2$ [n/s] | R [km] | Ref. |
|---|---|---|---|---|---|
| C/2010 U3 | 25.8 | | | | [1] |
| | 8.47 | 4.1 (28) | 3.5(27) | 10 | [2] |
| C/2014 UN271 | 23.8 | 2(28)* | | ~70 | [3] |
| | 20 | 7(27) | | | [4] |
| | 16.6 | 3.44 (27) | | | [5] |
| | 11 | 2(28) | | | [4] |
| C/2014 B1 | 9.7 | 8.5(27) | 9.9(28) | 6.4 | [2] |
| C/2017 K2 | 35 | | | 14 – 80 | [6] |
| | 23.7 | | | | [7] |
| | 16.4 | 1(28) | 8.9(26) | | [8] |
| | 15.8 | 1.6(28) | 1.4(27) | | [8] |
| | 12 | 4(28) | 3.4(27) | | [2] |
| C/2019 E3 | >20 | 4(26) | | >3 | [9] |
| C/2024 E1 | 7 | | 2.546(25) | <13.7 | [10] |
| Hale-Bopp | 6.7 | 2(28) | | 30 | [11] |

* Upper limit.

References: [1] Hui et al. (2019); [2] Milewski et al. (2024); [3] Farnham et al. (2021); [4] Lellouch et al. (2022); [5] Roth et al. (2025); [6] Jewitt et al. (2021); [7] Hui et al. (2017); [8] Meech et al. (2017); [9] Hui et al. (2024) [10] Snodgrass et al. (2025) [11] Biver et al. (1996)

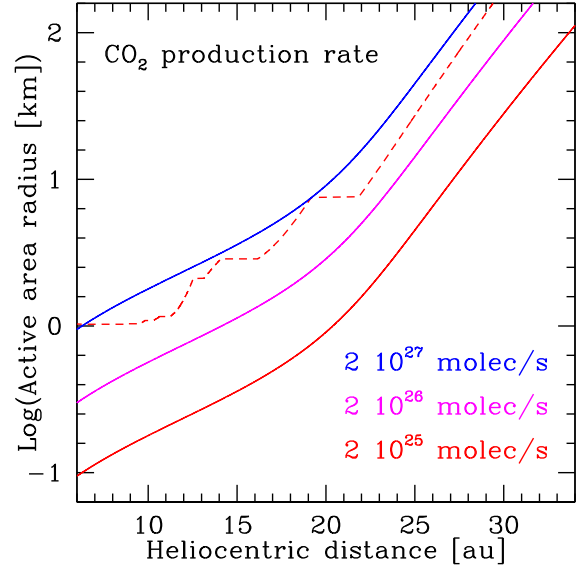

**Figure 12.** Theoretical estimate of the required radius of an active subsolar surface area producing $CO_2$ sublimation rates at detection limit (red curve), 10 times higher (purple curve), and 100 times higher (blue curve), as a function of heliocentric distance. The dashed curve shows the full model results for the subsolar area corresponding to the detection limit (see text). Relatively small active areas could suffice to produce detectable $CO_2$ activity even beyond 20 au.

For long-term evolution simulations focused on the comet's interior, this is a good approximation. However, the subsolar temperature is higher, and for particular configurations (surface geometry and spin-axis inclination), this temperature may be more relevant to the production rate of volatiles that lie close to the surface, like $CO_2$ in our case. The area exposed directly to the Sun will evaporate at a much higher rate, which means that the comet could be detected further away from the Sun.

To get an idea of the detection distance range, we use energy balance in a subsurface layer where $CO_2$ ice sublimates. With $Z(T)$ the sublimation rate and $H_{CO_2}$ the latent heat of $CO_2$ sublimation, we obtain the subsolar temperature $T_{SS}(d)$ as the solution of

$$\frac{(1-A)L_\odot}{4\pi d^2} = \epsilon \sigma T_{SS}^4 + Z(T_{SS})H_{CO_2}. \qquad (11)$$

We can now estimate the extent of a circular active area that would produce a desired production rate $\phi$ as a function of heliocentric distance by $\pi r_a^2 = \phi/Z(d)$, where $r_a$ is the radius of the area. The result is shown in Fig. 12 for the minimal production rate required for detection and two higher production rates.

A more realistic estimate of subsolar activity, albeit model-dependent, may be obtained by using the evolution code in the quasi-3D approximation (Prialnik et al. 2004), which is appropriate for a subsurface layer, considering only the solar flux at the subsolar point. We used this approach for the 50 km model in phase 4 (return from the OC) to obtain the $CO_2$ molecular flux as function of heliocentric distance, and based on it, the corresponding active area that would yield a production rate of $2 \times 10^{25}$ molecules $S^{-1}$. Here, the fraction of solar energy that goes into CO sublimation and conduction to the interior is taken into account as well. The required active area is therefore larger than the analytical estimate, but still acceptable.

Thus, due to the high sensitivity of the sublimation rate to temperature, we may still obtain $CO_2$ production rates above the detection limit at distances greater than that corresponding to the homogeneous surface temperature, even as far as 20 au and beyond.

### 5.3 Summary of evolution simulations and conclusions

We have followed the evolution of three objects of different sizes (radii of 2, 10 and 50 km)—initially composed of a mixture of dust, amorphous water ice, and CO and $CO_2$ ices—from formation to present time (4.6 Gyr), adopting a typical dynamical path: starting in the planetary region, close to Neptune's orbit, moving out gradually to the OC, where most of the time is spent, and returning to the planetary region on a highly eccentric orbit. Along with the dynamical evolution, full thermal and compositional evolution was followed, including sublimation/condensation of volatiles with the exchange of energy involved, and vapor flow through the porous bodies. The implicit numerical solution allowed time steps to vary by orders of magnitude, adjusting to the various timescales of the different processes governing the evolution.

The aim of this study was twofold: to explain in general terms the distant activity of OC comets and possibly to link the activity to size and past history. Our main conclusions, based on the single, albeit typical, orbit considered, are as follows.

- Small objects lose all their CO ice during the early evolutionary phase near the orbit of Neptune. At this stage, we can only state that the upper size limit for CO depletion is a diameter of 4 km.
- Large objects become CO-depleted to depths of order 500 m during early evolution, ~ 30 Myr in the Neptune region. Longer times will result in larger depths.
- For objects that retain CO in the deep interior, during the passage to the OC, we find that as the surface cools, the CO vapor migrates outward and refreezes close to the surface. It is this refrozen CO that will constitute the source of activity upon return from the OC. We note that a very similar effect, although on infinitely smaller scales (in both time and extent), namely cometary outbursts at sunrise, was observed by *Deep Impact* on 9P and by *Rosetta* on 67P: the cometary surface is heated at noon and the water ice front recedes from the surface; during the night, the surface cools and vapor produced at the ice front refreezes just below the surface; at sunrise this fresh ice evaporates, resulting in a short outburst (Prialnik et al. 2008; De Sanctis et al. 2015).

- Some of the deep CO ice may be further lost during the long sojourn in the cold OC, due to the heat released by the decay of long-lived radioactive elements. However, we have shown that this has only a minor effect on the activity at later stages. The effect of short-lived radioisotopes during early evolution is negligible for the cases considered here (see Parhi & Prialnik 2023).
- Since during early evolution, the temperatures remain too low for $CO_2$ to sublimate or the amorphous ice to crystallize, the $CO_2$ ice and the amorphous water ice are preserved in all objects considered until the return from the OC. Therefore, it is expected that some OC comets will be activated by $CO_2$ rather than CO.
- We find that on the inbound orbit from the OC: activity by CO emission starts around 50 au; activity by emission of $CO_2$ starts at ∼ 13 au and the release of trapped volatiles from crystallizing amorphous ice exceeds the other sources inward of 7 au.
- Finally, we note that nowhere along the new comet's orbit does the CO/$CO_2$ production rate ratio reflect the abundance ratio of these volatiles in the initial composition.

Encouraged by the success of the long-term evolutionary calculations presented here in elucidating many puzzling observations of distant comets originating in the OC, we shall proceed to investigate—in future studies—the role of crucial parameters such as the dynamical path, the cometary size and the initial composition, and their effect on the behavior of such comets.

## DATA AVAILABILITY

The data generated in this research will be shared on reasonable request to the corresponding author.